\documentclass[a4paper, 12pt]{article}
\usepackage{amsfonts}
\usepackage{amssymb}
\usepackage{latexsym}
\usepackage{enumerate}
\usepackage{amsmath}
\usepackage{amsthm}
\usepackage{graphicx}
\usepackage{multirow}
\usepackage{epstopdf}
\usepackage{epigraph} 
\usepackage{caption}
\usepackage[]{algorithm2e}
\DeclareGraphicsExtensions{.eps .pdf,.png,.jpg,.mps,.bmp}
\renewcommand{\baselinestretch}{1.2}

\begin{document}

\title{Optimal Policy Learning: \\ From Theory to Practice} 
\author{Giovanni Cerulli \\ {\small CNR-IRCRES} \\ {\small Research Institute on Sustainable Economic Growth} \\ 
{\small National Research Council of Italy} \\ {\small giovanni.cerulli@ircres.cnr.it}
\\ {\small \textbf{\textit{}}}}

\maketitle

\begin{abstract}
Following in the footsteps of the literature on \textit{empirical welfare maximization}, this paper wants to contribute by stressing the policymaker perspective via a practical illustration of an optimal policy assignment problem. More specifically, by focusing on the class of \textit{threshold--based} policies, we first set up the theoretical underpinnings of the policymaker selection problem, to then offer a practical solution to this problem via an empirical illustration using the popular LaLonde (1986) training program dataset. The paper proposes an implementation protocol for the optimal solution that is straightforward to apply and easy to program with standard statistical software.

\end{abstract}
\renewcommand{\baselinestretch}{1}
%

\newpage

\section{Introduction}
\label{sec_1}

Evidence-based program evaluation is increasingly becoming an essential tool for policymakers to fine--tune socio--economic policies (Cerulli, 2015). Generally, decision-makers run multiple rounds of similar policies at national, regional, or local level. As recurrent policies take place, they can collect massive evidence on past programs' operation and effects that increasingly enables a learning process based on past experience.

Program evaluation draws on three dimensions: ex--ante, in--itinere, and ex--post. Abstracting from in--itinere evaluation, this article focuses on how decision-makers can improve ex-ante evaluation of policy effects based on the results obtained from the (antecedent) ex-post evaluations. In tune with the recent literature, I call this process \textit{policy learning} (Athey and Wager, 2019).

How does this process take place? And more specifically, what do we mean by ex - ante program evaluation in this specific context? Answering these questions requires delving into a deeper understanding of the causal dynamic underlying a policymaker's action.

To fix the idea, consider a policymaker administrating a policy $P$ targeting a specific outcome $Y$. Either for ethical or policy constrains reasons, the policymaker hardly wants to (or can) pick up beneficiaries at random. She generally adopts selection strategies aimed at optimizing her objective function, typically by increasing the overall policy effect as much as possible. The point in question is: which effect?

It is well-known that the total effect (let's call it $\gamma$) of a policy intervention can be decomposed into two sub-effects, a direct one (let's call it $\alpha$), and an indirect one (let's call it $\beta$). The direct is the effect of the policy ``as if'' the decision--maker had selected both beneficiaries and non--beneficiaries at random. The indirect effect is that part of the total effect due to the selection process operated by the policymaker's choice of beneficiaries (the so--called \textit {policy assignment rule}). In this setting, we thus have:

\begin{equation}
\gamma(S) = \alpha + \beta(S)   
\end{equation}
where the indirect effect is clearly a function of the selection process $S$ operated by the policymaker. Different policy assignment rules affect the total effect (or \textit{welfare}) of the policy, thus raising a quest for an ``optimal'' policy rule within certain classes of feasible rules (Manski, 2004).

Ex--post program evaluation techniques, either experimental (randomized control trials, RCT) or quasi--experimental (re-balancing procedures), generally aim at consistently estimating $\alpha$ by assuming $\beta$ to be a bias induced by the selection process. This is the econometrician perspective, that does not coincide however with the policymaker viewpoint. Given budget, ethical, or contextual constraints, the policymaker is in fact mainly interested in increasing the total effect of the policy, and thus finding the selection rule $S^{*}$ which maximizes this effect (Dehejia, 2005).

Recently, the literature has produced significant theoretical advances on optimal policy assignment, both in statistics, biostatistics and econometrics\footnote{Specifically, this paper refers to the econometric papers of Manski (2004) and Kitagawa and Tetenov (2012) who extended the former. In his seminal paper, Manski (2004) assumes different probability distributions of the possible experimental outcomes, and considers treatment rules assigning individuals to those treatments producing the best experimental outcomes. These ``conditional empirical success'' rules consider different subsets of the observed covariates space. The author derives bounds for the welfare regret associated to such rules showing that it converges to zero at a rate of at least $0.5$. Kitagawa and Tetenov (2012) extend Manski approach to a broader class  of rules, including threshold-based ones, and show that similar rate of convergence are achieved by these broader class of rules. Other contributions to the literature are: Dehejia, 2005; Hirano and Porter, 2009; Athey and Wager, 2019; Bhattacharya and Dupas, 2012; Zhao et al., 2012; Zhang et al., 2012.}. Following in the footsteps of this literature, this paper wants to contribute by stressing the policymaker perspective via a practical implementation of an optimal policy assignment problem (or, equivalently, of an ``empirical welfare maximizing'' problem). More specifically, by focusing on the class of \textit{threshold--based} policies, we first set up the theoretical underpinnings of the policymaker selection problem, to then offer a practical solution to this problem via an empirical illustration using the popular LaLonde (1986) training program dataset. Moreover, the paper proposes an implementation protocol of the optimal solution that can be easily programmed with standard statistical software\footnote{With regard to the application presented in this paper, I developed a Stata 16 program for both a univariate and a bivariate empirical welfare optimization threshold--rule. All codes are available upon request.}.

The structure of the paper is as follows. Section \ref{sec_2} presents the theoretical underpinnings of the optimal constrained treatment assignment. Section \ref{sec_3} proposes a protocol for carrying out optimal constrained treatment assignment. Section \ref{sec_4} illustrates an empirical application using the popular LaLonde (1986) job training dataset and discusses the results. Section \ref{sec_5} concludes the paper.

\section{Optimal constrained treatment assignment}
\label{sec_2}

Let $X$ be an individual's vector of characteristics, $Y$ an outcome of interest, $T=\{0,1\}$ a binary treatment.
A policy assignment rule $\mathcal{G}$ is a function mapping $X$ to $T$, specifying which individuals are or are not to be treated:

\[\mathcal{G}: X  \rightarrow T \]
Define the policy conditional average treatment effect as:

\begin{equation*}
\tau(X)=E(Y_{1}|X) - E(Y_{0}|X)
\end{equation*}
where $Y_{1}$ and $Y_{0}$ represent the two potential outcomes of the policy, and $E_{X}[\tau(X)]=\tau$ the average treatment effect. 
The policy actual total effect (or \textit{welfare}) $W$ is defined as:

\begin{equation*}
W = \sum_{i=1}^{N}T_{i}\cdot\tau(X_{i})
\end{equation*}
and the policy \textit{unconstrained} optimal total effect (or \emph{unconstrained maximum welfare}) as:
\begin{equation*}
W^{*} = \sum_{i=1}^{N}T^{*}_{i}\cdot\tau(X_{i})
\end{equation*}
where:

\begin{equation*}
T^{*}_{i}=\mathbf{1}[\tau(X_{i})>0]
\end{equation*}
is the optimal unconstrained policy assignment.

The difference between the (unconstrained) maximum achievable welfare and the 
the welfare associated to the policy actually run is called \emph{regret}, and it is defined as:

\begin{equation*}
Regret = W^{*} - W
\end{equation*}
The regret is generally positive as decision makers are unable to cherry-pick the best treatment assignment possible, given their constrains and available information. 
This is particularly true for randomized control trials (RCT) as treatment is, in this case, deliberately void of any selection objectives. 

Table \ref{tab:tab1} sets out an illustrative example of the previous setting with three treated and three untreated units. The conditional average treatment effect $\tau(X)$ is reported in the third column of the table, while the last row of the fourth column shows the actual policy welfare that in this numerical example is equal to $10$. By definition, the optimal (unconstrained) treatment assignment would select only those units obtaining a positive $\tau(X)$, that in this specific case are the units 1, 3, 4, and 6. This is the optimal selection assignment $T^{*}$, with the last row of the last column showing that the largest welfare achievable by the policymaker is equal to $26$. The regret is therefore equal to $26-10=16$. 

Because of eligibility, budget, ethical, or institutional constrains, policymakers are in general unable to implement the optimal unconstrained policy assignment,
being therefore obliged to rely on a \emph{constrained} assignment $T'$  which selects treated units according to (some) of their characteristics.
The welfare thus obtained, $W'$, may or may not be lower than $W^{*}$. However, one can ask whether it is possible for policymakers to produce the largest possible constrained welfare.
As the set of feasible treatment assignment rules is huge, replying to the previous question requires to restrict the focus on certain \textit{classes} of policies and find, among them, the optimal assignment to treatment (\textit{optimal constrained policy assignment}). 

Although there are several classes of policies, policymakers often use few of them, such as ``threshold-based'', ``linear--combination'', or ``fixed-depth decision trees'' (Kitagawa and Tetenov, 2018).

Threshold-based assignment policies are popular in policy programs as they are simple to manage and easy to interpret. For this class of policies, policymakers select one or more than one variable of outstanding importance (such as, for example, firm number of employees, bank total asset, individual age, etc.) by selecting specific values (the thresholds) discriminating between treated and untreated units. To clarify, suppose to have just one single selection variable $x$, with threshold $c$. The assignment to treatment is clearly a function of $c$:

\begin{equation*}
T_{i}(c)=T^{*}_{i} \cdot \mathbf{1}[x>=c]
\end{equation*}
with corresponding welfare:
\begin{equation*}
W(c) = \sum_{i=1}^{N}T_{i}(c)\cdot\tau(X_{i})
\end{equation*}
We define the optimal choice of the threshold $c$ as the one maximizing $W(c)$ over $c$:

\begin{equation*}
c^{*}=\mathtt{argmax}_{c}[W(c)] 
\end{equation*}
If $c^{*}$ exists, the optimal constrained welfare will thus be equal to $W(c^{*})$. 

Extensions to the case of two or three selection variables are straightforward. Suppose to have two selection variables, $x$ and $z$, and two corresponding thresholds, $c_{x}$ and $c_{z}$. In this case, the assignment to treatment is a function of both thresholds, and takes on this form:

\begin{equation*}
T_{i}(c_{x},c_{z})=T^{*}_{i} \cdot \mathbf{1}[x>=c_{x}] \cdot \mathbf{1}[z>=c_{z}]
\end{equation*}
\\  
The previous welfare, $W(c)$, can thus be maximized over $c$ as a vector of thresholds, one for each selection variable. In two dimensions, the previous assignment rule is called \textit{quadrant--assignment}, as it selects the upper--right quadrant of the four quadrants generated by setting the thresholds. 

There are some problematic aspects that can arise when seeking the optimal thresholds. It is possible that, at the optimal thresholds, the share of the units to treat could be too small, or specularly, too large. In this case, it would be meaningless to run a policy based on such too small (or too large) number of treated. To solve the problem, the policymaker can however consider either budget constraints (for example, the maximum number of units she would be able to treat given the availability of a certain amount of money), or set in advance a targeted number of units to treat (for example, a treatment share between $30\%$ and $40\%$ of the entire reference population). This procedure might lead to a sizable reduction of the welfare but would preserve the full feasibility of the policy.       

A related problem arises when, for one or more selection variables, there exists the so-called \textit{angle solution} due to a monotonic effect of the selection variable(s) on the welfare. To give a practical example of this occurrence, consider as selection variable the educational attainment of individuals. In many policy contexts, the welfare (or effect) of the policy increases monotonically with the level of education. This event would lead to select only people having the highest level of educational attainment in the sample. This can be however unfeasible for two reasons: (i) the policy aims at targeting poorly educated people; (ii) the number of treated units would become too small as people with high educational attainment are generally few. The solution to this problem is similar to the one set out above, that is, the policymaker can again introduce a threshold limit, and/or a prefixed range of treatment shares, and thus run the policy according to these constraints. Again, this procedure may entail a shrinking welfare, but would preserve the possibility to eventually run the policy.   

\begin{table}[hbt]
\centering
$\begin{array}{*{20}{c}}
\hline
\; \; \; \; \; \;  {ID} \; \; \; \; \; \;  &  \; \; \; \; \; \;  {T} \; \; \; \;   \; \; & \; \; \; \; \; \;   {\tau(X)} \; \; \; \;   \; \; & \; \; \; \; \; \;   {T \cdot \tau(X)} \; \; \; \;   \; \; &  \; \; \; \; \; \;   {T^{*}} \; \; \; \;  \; \;  & \; \; \; \; \; \;    {T^{*}\cdot \tau(X)}  \; \; \; \;   \; \; \\
\hline
1 & 1 & 9  & 9 & 1  & 9\\
2 & 1 & -4 & -4  & 0  & 0 \\
3 & 1 & 5  & 5  & 1  & 5 \\
4 & 0 & 6  & 0  & 1  & 6 \\
5 & 0 & -2  & 0 & 0  & 0 \\
6 & 0 & 6 & 0  & 1  & 6 \\
\hline
 &  &  & 10  &   & 26 \\
\hline
\end{array}$
\caption{Example of an optimal policy assignment rule. The regret of this policy is equal to $16=26-10$.}
\label{tab:tab1}
\end{table}

\begin{figure}[ht]
\centering
\includegraphics[width=12cm]{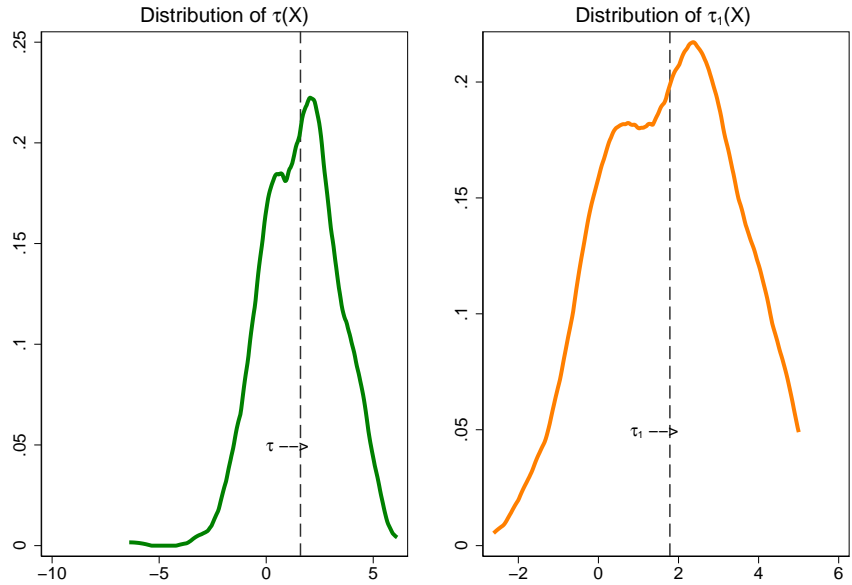}
\caption{Distribution of $\tau$(X) and $\tau_{1}$(X). Program: National Supported Work Demonstration (NSWD). Data: LaLonde (1986). Target variable: Real earnings in 1978. Estimation technique: Regression--adjustment (with observable heterogeneity).}
\label{fig:fig1}
\end{figure}

\section{The empirical welfare maximizing protocol}
\label{sec_3}

To provide an empirical strategy for carrying out a threshold-based optimal policy assignment decision,
we suggest the following procedure consisting of five steps. 

\vspace{0.5cm}

\parbox{\textwidth}{

\noindent\rule{15.5cm}{0.4pt} \newline
\textbf{Procedure}. Threshold--based optimal policy assignment  \newline
\vspace{-1cm}

\noindent\rule{15.5cm}{0.4pt} \newline

\vspace{-1cm}
\begin{enumerate}

\item Suppose to have data from an RCT or from an observational study consisting of the information triple ($Y, X, T$) available for every unit involved in the program. 

\item Run a quasi--experimental method with observable heterogeneity, estimate $\tau(X)$, and compute the actual total welfare of the policy $W$.

\item Identify the optimal unconstrained policy $T^{*}$, and compute $W^{*}$, i.e. the maximum total welfare achievable by the policy, and estimate the regret as $W^{*} - W $.

\item Consider a constrained selection rule $T(x,c)$ based on a given set of selection variables, $x$, and related thresholds, $c$, and define the maximum constrained welfare as $W(x,c)$.

\item Build a greed of $K$ possible values for $c \in \{c_{1},...,c_{K} \}$, compute the optimal vector of thresholds $c_{k^{*}}$ and the corresponding maximum welfare $W(x,c_{k^{*}})$ thus achieved. 

\end{enumerate}   
\vspace{-0.5cm}           
\noindent\rule{15.5cm}{0.4pt} \newline
}

In the application proposed in the next section, we will show how to implement this procedure on a real policy dataset. 

\section{Application}
\label{sec_4}

We consider the popular experimental dataset from the National Supported Work Demonstration (NSWD) used by LaLonde (1986).  As well-known, this study looked at the effectiveness of a job training program (the treatment) administrated in 1976 on the real earnings of individuals two years after the completion of the program. It includes a set of demographic, social and economic variables at individual level such as age, race, educational attainment, previous employment condition, and real earnings - as well as the treatment indicator, and the real earnings in 1978, that is our target variable.

We first estimate the average treatment effect of this program using a regression--adjustment (RA) approach and a specification of the model including as control variables: real earnings in 1974 and 1975, age, age squared, an indicator for not having a degree, an indicator for being married, one for being black, and one for being hispanic.  

As the assignment to treatment of this program was randomized, the average treatment effect (ATE) is consistently estimated by the treated--control's difference--in--mean (DIM), found to be equal to 1.79 thousand dollars. For the sake of comparison, however, we also estimate ATE by a regression--adjustment obtaining a similar estimation value, 1.76, still significant at 1\%. The RA approach allows us to estimate also the average treatment effects conditional on the covariates, i.e. $\tau(X)$ and $\tau_{1}(X)$, as well as their distributions as set out in Figure \ref{fig:fig1}.

\begin{figure}[t]
\centering
\includegraphics[width=15cm]{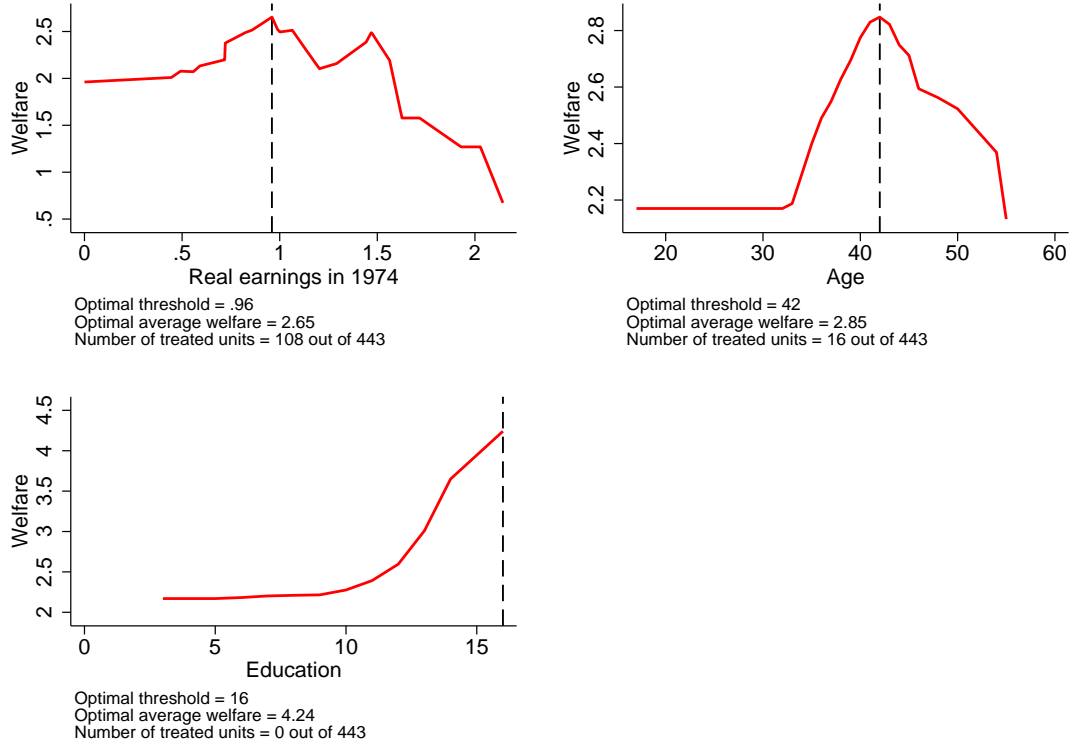}
\caption{Computation of the policy optimal selection threshold in univariate cases. Program: National Supported Work Demonstration (NSWD). Data: LaLonde (1986). Target variable: real earnings in 1978. Univariate selection variables: real earnings in 1974, age, and educational attainment.}
\label{fig:fig2}
\end{figure}

\begin{figure}[t]
\centering
\includegraphics[width=15cm]{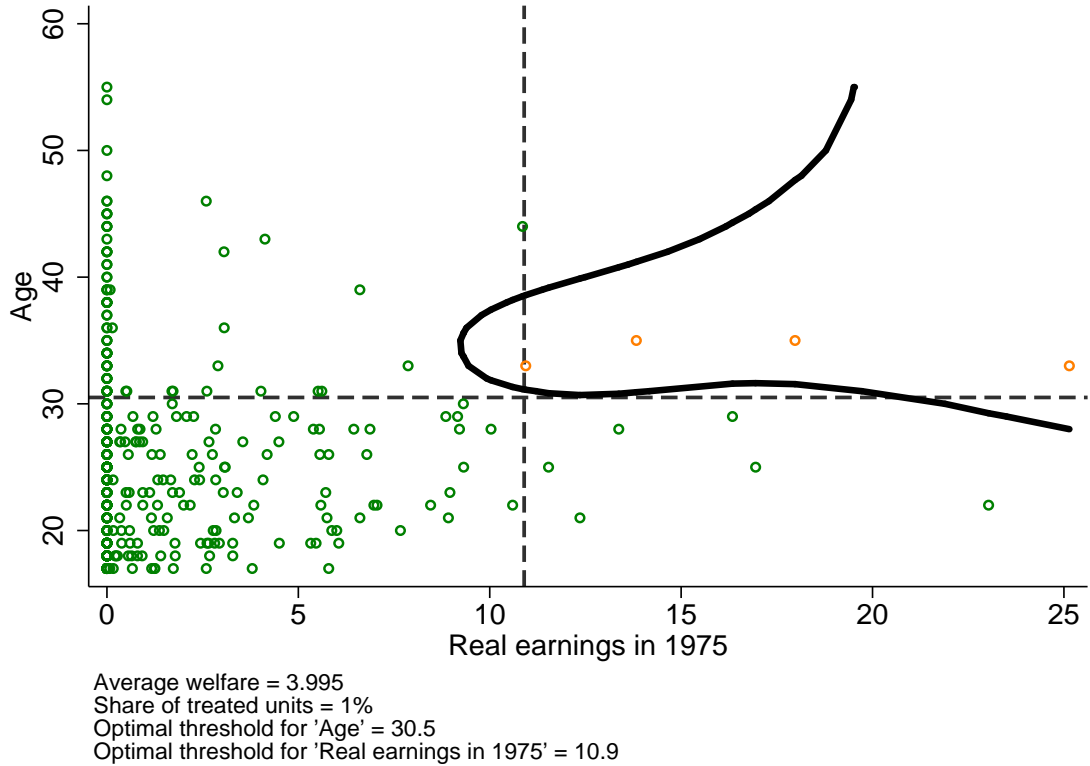}
\caption{Computation of the policy optimal decision boundary in the bivariate case. Program: National Supported Work Demonstration (NSWD). Data: LaLonde (1986). Target variable: real earnings in 1978. Bivariate selection variables: real earnings in 1975 and age.}
\label{fig:fig3}
\end{figure}

\begin{figure}[ht]
\centering
\includegraphics[width=15cm]{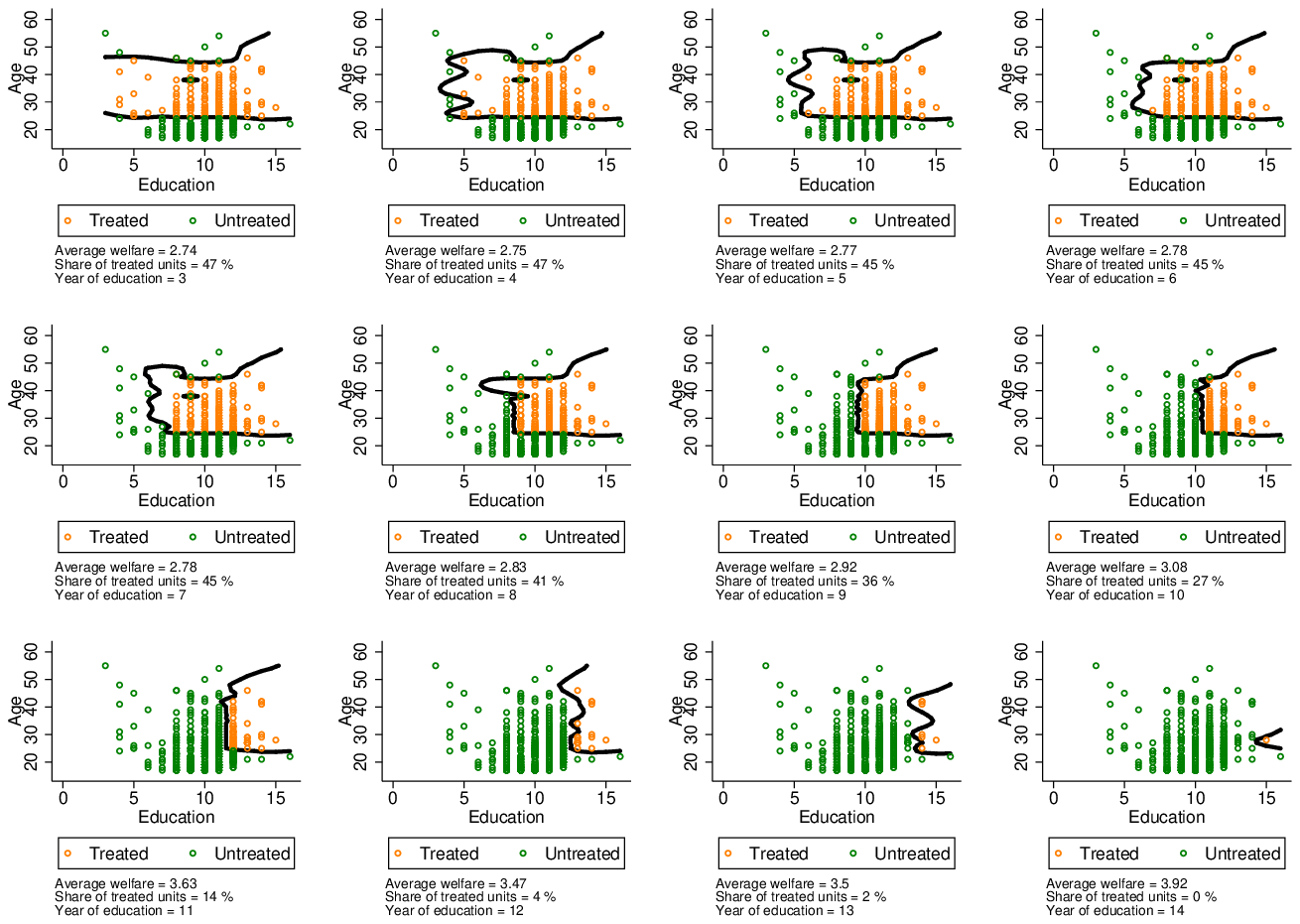}
\caption{Computation of policy optimal decision boundaries in the bivariate case, when one of the two selection variables (age) is fixed at its optimal threshold, and the threshold of the other variable (education) is varying. Program: National Supported Work Demonstration (NSWD). Data: LaLonde (1986). Target variable: real earnings in 1978. Bivariate selection variables: age and educational attainment.}
\label{fig:fig4}
\end{figure}

Now we move on by implementing the previous empirical welfare maximizing protocol, first in a univariate, and then in a bivariate selection setting. 

The univariate setting considers only one selection variable. For the sake of comparison, we run the empirical welfare maximizing protocol separately on four selection variables, i.e. real earning in 1974, age, and education. For every selection variable, we compute the optimal threshold, namely, the selection criterion that would allow the policymaker to maximize the average welfare. Figure \ref{fig:fig2} illustrates the results. 

The graphs herein reported show that, for real earning in 1974, the optimal threshold is $0.96$, with corresponding maximum average welfare equal to $2.65$. Results for age show that the welfare maximizing threshold is found at age 42, with a corresponding average welfare of $2.85$. This means that, if the policymaker had selected as beneficiaries people aged 42 or larger, she would have been able to obtain an increase of the welfare by 1.09 thousands dollars per beneficiary per year, which is the difference between the pure RCT effect, 1.79, and the one obtained by selecting over age. 

Results for education are particularly interesting as in this case we have an angle solution due to monotonicity. It means that the empirical welfare maximization of this program would lead to select people with the highest educational attainment possible (which is around 15 years). This is clearly unfeasible, as no one would be selected in this case. As said above, this is however a minor issue, as we could adjust our empirical welfare maximization by requiring a certain ``minimum'' share of units to treat, so to make our analysis eventually feasible. Of course, adding this constraint would lead to a reduction of the obtained level of the welfare (that, in this specific case, is rather high and equal to 4.24). As the size of the welfare is huge, we expect that constraining on a minimum share of treated units can allow to obtain a larger welfare than the one obtained by randomizing the assignment to treatment.     

To explore this latter aspect, i.e. to what extent introducing a constrain on the share of treated would affect the average welfare, we consider a bivariate setting, i.e. a setting characterized by two selection variables. Before moving to this, however, it seems useful to first provide an example of a bivariate optimal treatment assignment. 

Figure \ref{fig:fig3} sets out an example where we carried out an empirical welfare maximization of the NSWD jointly over age and real earnings in 1975 (with real earnings in 1978 chosen again as target variable). The figure also shows the optimal estimated decision boundary\footnote{This is an estimate of the so--called \textit{Bayesian decision boundary} as defined by supervised learning classification models (Gareth et al., 2013). In our case, this boundary should be closer to the boundary of the upper-right quadrant: it is however smoother and imprecise due to the high sparseness (few observations) located in this quadrant, with the largest part of individuals placed in the lower--left quadrant.} (the curve drawn in black). Observe that in the univariate case treated above, the decision boundary collapses to a single point. The optimal thresholds are 30.5 for age, and 10.9 for real earnings in 1975. 

The upper-right quadrant represents the optimal treatment zone, and it is immediate to see that only four individuals are selected as beneficiaries. Unfortunately, they represent only the $1\%$ of the entire sample. A solution to this problem would be to provide the policymaker with a \textit{menu} of optimal treatment scenarios among which she can make pondered decisions. We show this meanu using as selction variables age and education within the same dataset. 

Figure \ref{fig:fig4} sets out the optimal policy assignment decision boundary over age at different education thresholds\footnote{The choice of varying the threshold of education by letting age fixed to its optimal threshold is dictated by education monotonicity as discussed above. It goes without saying that one might do the opposite as well.}. As said, this figure can be interpreted as a menu for the policymaker to choose among alternative scenarios characterized by different policy settings entailing a trade--off between the size of the policy effect and the number of units to treat. As long as the educational attainment threshold increases, we observe in fact that the average welfare increases too. The maximum of the average welfare is obtained when only one individual is treated, with an average welfare of 3.92 and an education of 14 years. Any other intermediate scenario of this menu would entail a smaller average welfare with, however, a higher number of treated units. Given a budget constrain, for example, the policymaker can cherry--pick one of these scenarios and run the policy using this menu as reference for an ex-ante optimal re-programming of the policy treatment assignment.

\section{Conclusion}
\label{sec_5}

The literature on empirical welfare maximization is growing up rapidly. Large availability of datasets on programs already carried out, either based on observational data or randomized control trials, allows researchers and policymakers to design policies for increasing social welfare by optimally fine-tuning treatment assignment. Following in the footsteps of this recent literature, this paper has stressed the policymaker's perspective by proposing a practical implementation of an optimal policy assignment problem within the class of threshold-based selection rules. Straightforward to apply in practice and to implement with standard statistical software, the proposed procedure and illustrative application can guide policymakers to improve the ex--ante design of future policies; in other words, to learn from experience.

\pagebreak

\end{document}